\documentclass[letterpaper, preprint, paper,11pt]{AAS}	% for preprint proceedings

\usepackage{bm}
\usepackage{amsmath}
\usepackage{comment}
\usepackage{textcomp}

\usepackage[colorlinks=true, pdfstartview=FitV, linkcolor=black, citecolor= black, urlcolor= black]{hyperref}
\usepackage{overcite}
\usepackage{footnpag}			      	% make footnote symbols restart on each page

\usepackage{graphicx}
\usepackage{epstopdf}
\AppendGraphicsExtensions{.tif}
\usepackage{algorithm}
\usepackage{algpseudocode}
\usepackage{pgfgantt}
\usepackage{svg}
\usepackage{multirow}
\usepackage{subcaption}

\PaperNumber{24-145}

\begin{document}

\title{Pose estimation of CubeSats via sensor fusion and Error-State Extended Kalman Filter}

\author{Deep Parikh\thanks{PhD Student, Aerospace Engineering, Texas A\&M University, College Station, TX.}
\ and 
Manoranjan Majji\thanks{Associate Professor, Aerospace Engineering, Texas A\&M University, College Station, TX.}
}

\maketitle{} 		

\begin{abstract}A pose estimation technique based on error-state extended Kalman that fuses angular rates, accelerations, and relative range measurements is presented in this paper. An unconstrained dynamic model with kinematic coupling for a thrust-capable satellite is considered for the state propagation, and a pragmatic measurement model of the rate gyroscope, accelerometer, and an ultra-wideband radio are leveraged for the measurement update. The error-state extended Kalman filter framework is formulated for pose estimation, and its performance has been analyzed via several simulation scenarios. An application of the pose estimator for proximity operations and scaffolding formation of CubeSat deputies relative to their mother-ship is outlined. Finally, the performance of the error-state extended Kalman filter is demonstrated using experimental analysis consisting of a 3-DOF thrust cable satellite mock-up, rate gyroscope, accelerometer, and ultra-wideband radar modules.
\end{abstract}

\section{Introduction}
\begin{figure}[tb]
     \begin{subfigure}[b]{0.4\textwidth}
        \centering
         \includegraphics[width=\textwidth]{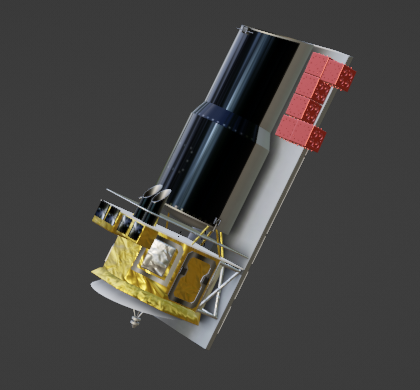}
     \end{subfigure}
     \centering
     \begin{subfigure}[b]{0.37\textwidth}
        \centering
         \includegraphics[width=\textwidth]{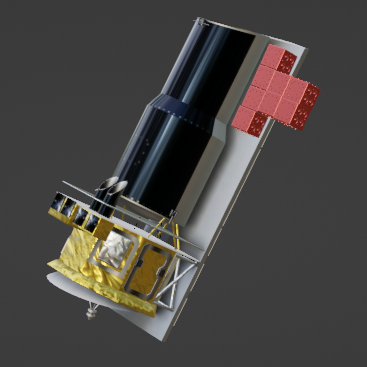}
    \end{subfigure}
    \caption{Various scaffolding structures to enable servicing of RSO}
    \label{fig:scafolding}
\end{figure}

The Land, Air and Space Robotics (LASR) laboratory at the Texas A\&M University is investigating the feasibility of employing thrust-capable Transforming Proximity Operations and Docking Service (TPODS) CubeSat modules to gain custody of Resident Space Objects (RSO)\cite{TPODS_detumble}. The proposed solution consists of a mother-ship carrying a few TPODS modules that approaches the RSO, analyzes the tumbling motion, and deploys TPODS modules towards the RSO\cite{TPODS_system}. Once a sufficient number of TPODS are attached to the RSO, the on-board thruster of TPODS can be first utilized to estimate the inertial properties of the RSO\cite{TPODS_estm} and then to apply relevant forces to regain the stable attitude. At this stage, TPODS modules involved in the detumbilng can be further employed to create a scaffolding structure as shown in Figure~\ref{fig:scafolding}\footnote{NASA's Spitzer Space Telescope by NASA Visualization Technology Applications and Development (VTAD)} to allow future servicing vehicles to dock with the RSO. The red-colored cuboid structure represents TPODS CubeSats modules in Figure~\ref{fig:scafolding}.

During the scaffolding generation phase, accurate pose estimation of each TPODS module is required due to the involvement of precise proximity operations and docking. Considering the proposed form-factor of 1U for TPODS modules, position accuracy in the range of centimeters and orientation accuracy of a few degrees is needed to ensure reliable docking of TPODS for scaffolding generation\cite{TPODS_ICRA}. Most conventional approaches to achieve accurate positional awareness for small satellite proximity operations rely on using multiple Global Positing System (GPS) receivers\cite{doi:10.2514/1.A35598}. With multiple GPS measurements, a highly accurate position on the satellite can be inferred. However, such solutions are plagued with high costs and restricted operating conditions. Under substantial tumbling conditions, GPS receivers on board the satellite are prone to erratic measurements\cite{Montenbruck2008}.

Each TPODS module consists of an Inertial Measurement Unit (IMU) that provides information about the angular rates and acceleration of the module. In addition to that, the module also consists of monocular vision as well as Ultra Wide Band(UWB) range measurement sensors. The proposed method fuses measurements from these sensors to estimate the pose with high accuracy using an error-state extended Kalman filter (EKF). The pose estimator performance is further validated for a pure translation, combined translation, and rotational motion scenarios. Finally, the algorithm is experimentally validated on a 3-DOF testbed, and the performance of the error-state EKF-based pose estimator is discussed.

\section{System Dynamics}
Once deployed from the mothership, each TPODS module can determine the relative range from the UWB anchors mounted on the booms attached to the ship. The extended boom structure enables range measurements and subsequent position triangulation with less dilution of precision\cite{langley1999dilution}. However, due to the geometry of the anchor placements, the dilution of precision will be comparatively poor in the z direction than in the $xy$ plane as the separation distance of anchors in the z direction is lower when compared to the $xy$ plane. Each TPODS module can move under the influence of the initial launch force and its thrusters. The motion governing equations of the TPODS are

\begin{equation}
m\boldsymbol{\Ddot{x}} = \mathbf{R}[f_xe_1+f_ye_2+f_ze_3] \label{eq}
\end{equation}
\begin{equation}
\mathbf{\dot{R}} = \mathbf{R}[\![\boldsymbol{\omega}\times]\!]
\end{equation}

Where the position of the TPODS module with respect to the inertial frame is denoted by $\boldsymbol{x}$ and the body's orientation with respect to the inertial frame is given by $\boldsymbol{R}$. The total external force exerted by the onboard thrusters in body-fixed $x$,$y$, and $z$ axis is given by $f_x$,$f_y$, and $f_z$, respectively. Further, $\boldsymbol{\omega}=\omega_1e_1 + \omega_2e_2 + \omega_3e_3$ is the angular velocity of the module expressed in the body frame and $[\![\boldsymbol{\omega}\times]\!]$ represents the standard cross product. It is assumed that planetary gravity affects both the mothership and TPODS equally.

\begin{figure}[tb]
     \begin{subfigure}[b]{0.59\textwidth}
	\centering
        \includegraphics[width=0.9\textwidth]{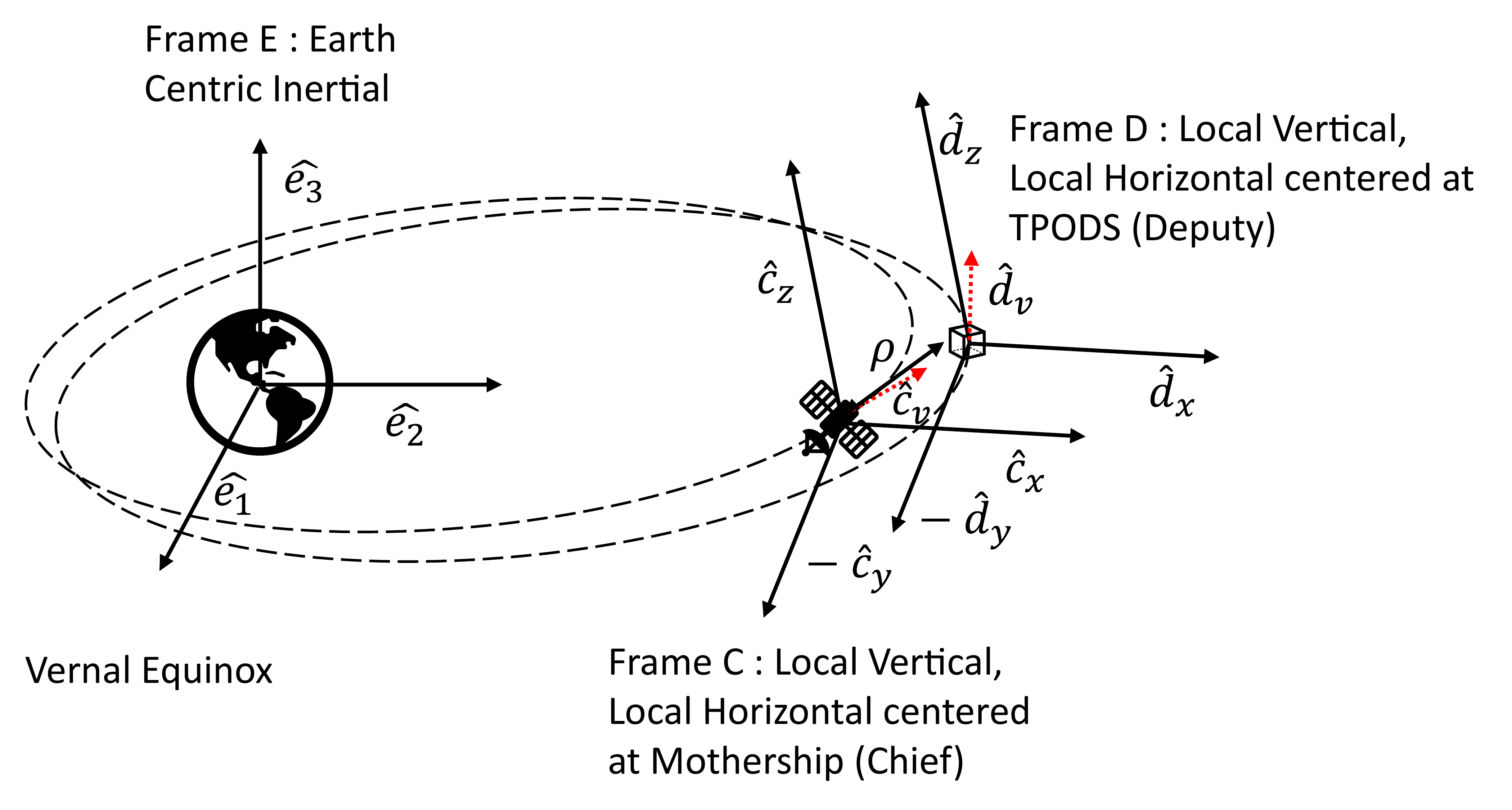}
	\caption{}
	\label{fig:frames}
    \end{subfigure}
	\centering
	\begin{subfigure}[b]{0.4\textwidth}
        \centering
	    \includegraphics[width=\textwidth]{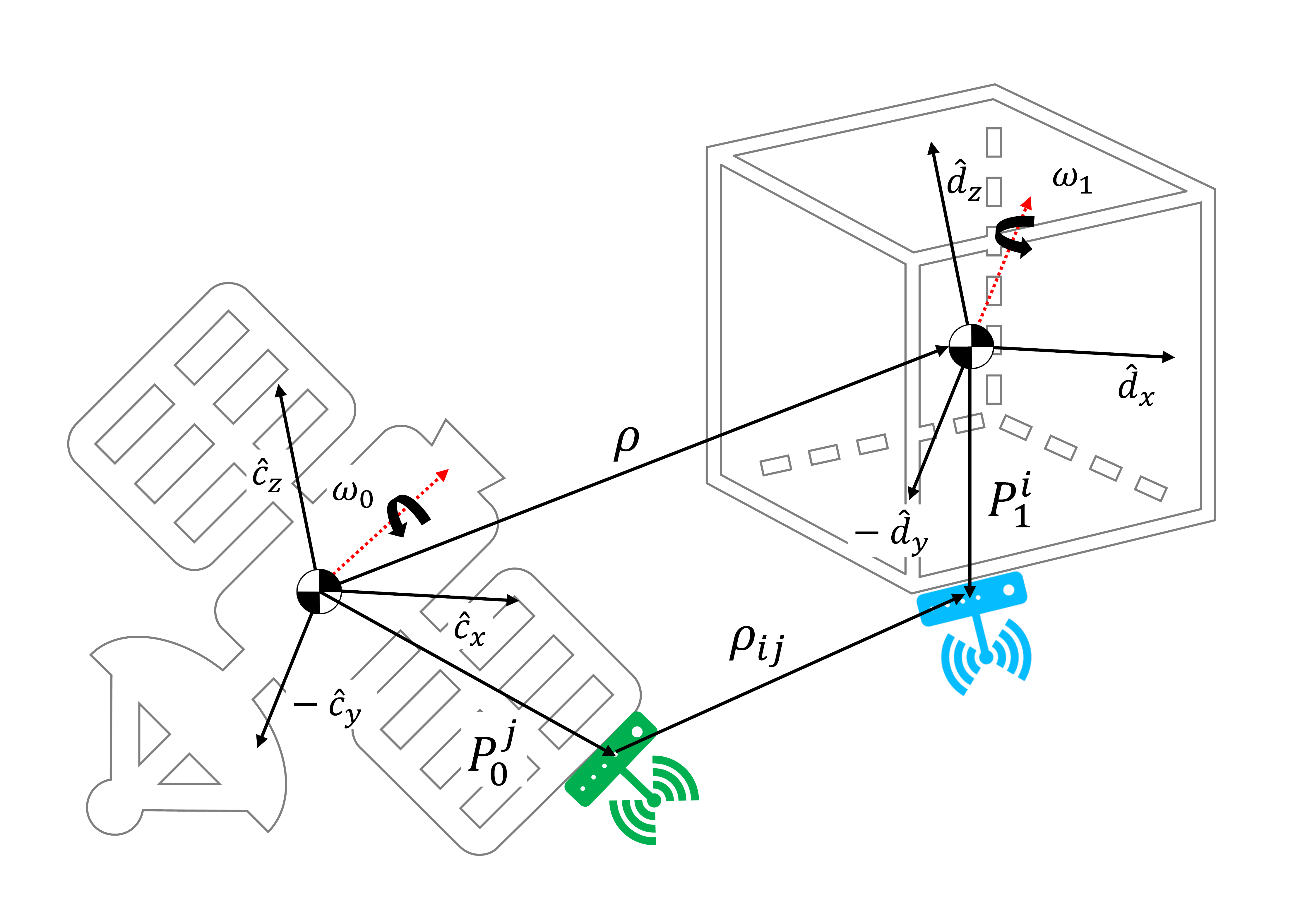}
	    \caption{}
	    \label{fig:coupling}
     \end{subfigure}
	\caption{Arrangement of UWB transceivers and frames of reference}
\end{figure}

\subsection{Traditional Approach using HCW Equations}
Traditional approaches outline a process to analyze the on-orbit relative motion of a deputy spacecraft relative to a chief, using Clohessy–Wiltshire (CW) equations\cite{HCW,Hill}. However, one of the fundamental assumptions of this analysis is the independence of the translation and rotational degrees of freedom. The equations of motion given by such analysis only provide the nature of the relative translation motion of a deputy spacecraft relative to a chief. Such equations can be further employed along with Gauss’ variational equations of motion to develop feedback control laws to achieve the desired relative orbit of both spacecraft\cite{battin1987introduction}. Such techniques have been successfully adapted to practical applications involving spacecraft formation flying for various observation missions\cite{naasz2002application}. However, such an approach might not render accurate for the application of spacecraft rendezvous for the reasons mentioned in the following discussion.

\subsection{Kinematic Coupling and Relative Motion}
As shown in Figure~\ref{fig:frames} several inertial and non-inertial reference frames can be considered to analyze the full 6-DOF motion of the deputy relative to the chief. The earth central inertial frame $\boldsymbol{e}$ serves as an inertial reference frame for the analysis. Two local vertical, local horizontal(LVLH) reference frames $\boldsymbol{c}$ and $\boldsymbol{d}$, centered at the chief and deputy, respectively, aid in analyzing the relative motion between the chief and deputy. The position vector of the deputy, relative to the chief, is $\boldsymbol{\rho}$, albeit it is the relative position vector between respective centers of masses. However, Figure~\ref{fig:coupling} shows the actual placements of the UWB transceivers on the chief and deputy. For practical applications, the UWB transceivers will be mounted with some offsets at the respective center of mass. Hence, it is vital to incorporate the effects of this offset in all dynamical equations during the estimation and closed-loop control design process.

\begin{figure}[tb!]
\centerline{\includegraphics[width=0.9\textwidth]{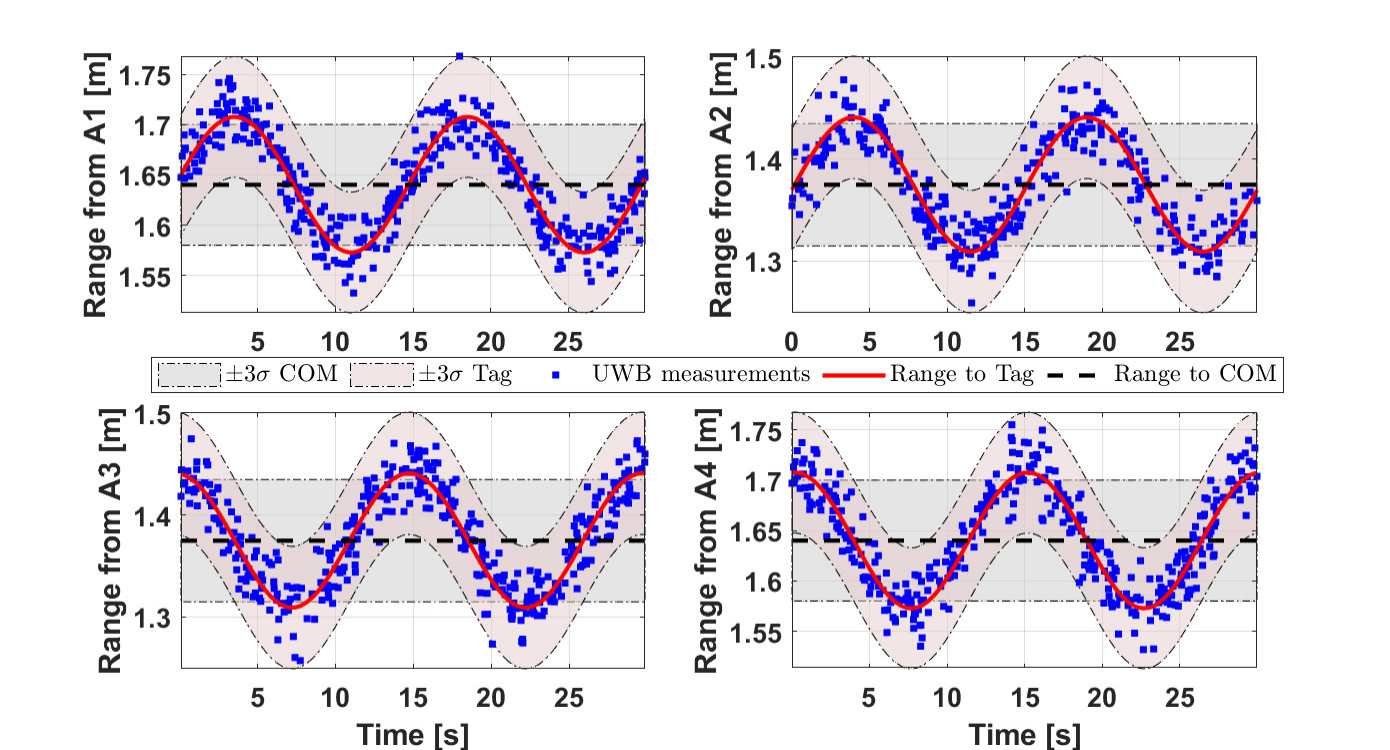}}
\caption{Anticipated measurements for rotating deputy}
\label{fig:meas_comap}
\end{figure}

Figure~\ref{fig:meas_comap} highlights one such scenario to explain the effect of kinematic coupling due to offset between the center of mass and the range sensor. For this simplified scenario, the chief spacecraft is assumed to be stationary, having four UWB anchors. The respective centers of mass of the chief and deputy spacecraft are one meter apart, and the UWB tag is mounted at $\mathbf{7.07cm}$ from the center of the deputy. When the deputy undergoes only rotational motion, the distance between both centers of mass remains constant. However, the range measured by the tag to various anchors shows an oscillatory trend due to the offset. As discussed in later sections on sensor models and pose estimators, the estimation routine expects the range measurements to be within certain bounds based on the sensor characteristics. If the dynamical relation between only respective centers of mass is provided to the estimation algorithm, it expects the range measurements within the constant grey bounds depicted in Figure~\ref{fig:meas_comap}. However, in reality, the range measurements follows the oscillatory red curve and stay within the red bounds. Incorporating these out-of-model measurements in the estimation process requires a higher amount of process noise to be injected, significantly degrading the quality of the pose estimate. 

Considering the fact that both range sensors are fixed to their respective rigid spacecraft bodies, the velocity and acceleration level dynamics of the relative range is governed by\cite{ALFRIEND2010227},
\begin{equation}
\boldsymbol{\dot{\rho}_{ij}} = \boldsymbol{\dot{\rho}}  + \boldsymbol{\omega} \times \boldsymbol{P_1^i}
\label{eqn:coupled_EOM} 
\end{equation}
\begin{equation}
\boldsymbol{\ddot{\rho}_{ij}} = \boldsymbol{\ddot{\rho}}  + \boldsymbol{\dot{\omega}} \times \boldsymbol{P_1^i} + \boldsymbol{\omega} \times \left(\boldsymbol{\omega} \times \boldsymbol{P_1^i}\right)  
\end{equation}

\section{Sensor Models}
\subsection{Rate Gyroscope and Accelerometer}
It is assumed that the rate gyroscope exhibits a well-studied behavior of Farrenkopf’s gyro dynamics error model, i.e. having a zero mean white noise as well as the bias governed by random walk \cite{Farrenkopf}. 
\begin{equation}
\boldsymbol{z}_{gyro} = \boldsymbol{\omega} + \boldsymbol{\eta}_{gyror} + \boldsymbol{b}_{gyro}
\label{eqn:gyro_sen}
\end{equation}
\begin{equation}
\boldsymbol{\Dot{b}}_{gyro} = \boldsymbol{\eta}_{gyrob}
\end{equation}
Where $\boldsymbol{\eta}_{gyror}$ depicts zero mean white noise in the angular rate measurements and $\boldsymbol{\eta}_{gyrob}$ represents the zero mean white noise that governs the random walk of the bias in angular rate measurements. However, since the duration of TPODS proximity operations is relatively small, the bias parameters are assumed to not change during the process. The sensor suite on the mothership spacecraft can be leveraged to set each TPODS's biases just before deployment towards the RSO. Hence, for the pose estimation and closed-loop control of TPODS, a simplified gyro model of Equation~\eqref{eqn:gyro_sen} is considered.

\subsection{UWB Radar} \label{sec:UWB_sen}
\begin{figure}[htbp]
\centerline{\includegraphics[width=0.7\textwidth]{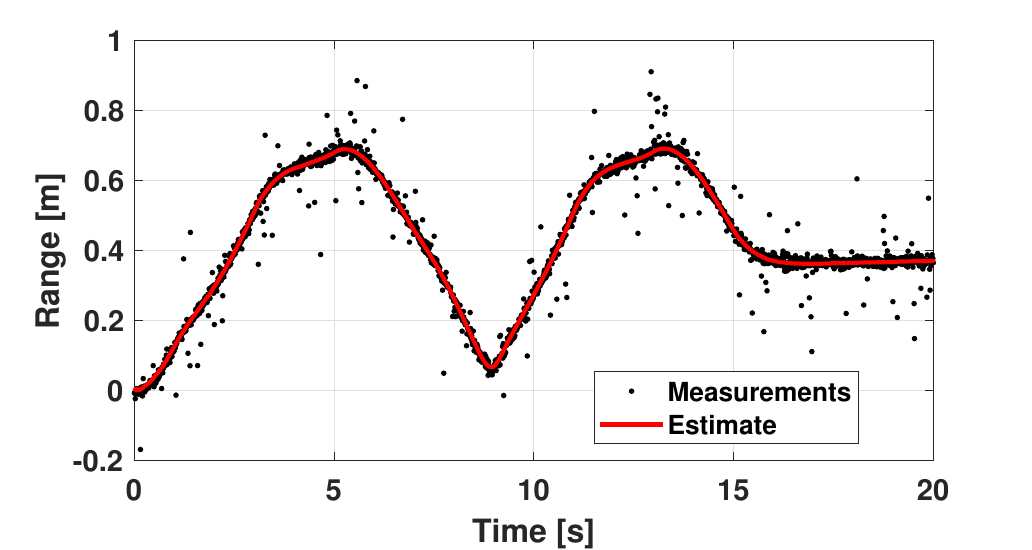}}
\caption{Simulated UWB range measurements and estimated range from EKF}
\label{fig:range_meas}
\end{figure}

The UWB radio modules inside the TPODS use the Time Of Flight (TOF) based range measurement technique to compute the relative distance between the module and fixed anchors\cite{mymanual}. However, the UWB range measurements are adversely affected by multi-path reflection and clock drift errors. Consequently, the range measurements often contain outliers and can significantly affect the accuracy of an estimated pose\cite{zhao2022uwbData}. Figure~\ref{fig:range_meas} presents a scenario where simulated range measurements are corrupted with measurement noise and outliers. The module is commanded to move under constant acceleration alternatively in X and Y directions. The range measurements are sampled from a Gaussian distribution with a true range as mean and standard deviation of 1 cm. For the outlier generation, $10\%$ of the total samples are randomly chosen and re-sampled from a Gaussian distribution with a standard deviation of 10 cm.

\section{Pose Estimator}
The main goal of the pose estimation algorithm is to accurately predict the relative position, velocity, and attitude of the TPODS module. To ensure that the estimator is robust from any adverse effects of the singular direction cosine matrix, an attitude quaternion depicts the module's attitude, and the pose estimation algorithm is formulated to embed the attitude quaternion. Furthermore, to avoid the growth of the computational cost associated with a total number of states in EKF, total states are divided into two tandem estimation routines\cite{Markley}. The angular rates and respective biases can be estimated with a separate EKF, while the remaining states, i.e., inertial position, body velocity, and attitude quaternion estimates, can be computed with an additional EKF. However, since the biases are assumed to be constant, only one estimator based on EKF is needed to predict position, velocity, and attitude quaternion.

Consequently, the estimated angular velocity can be written as
\begin{equation}
\hat{\boldsymbol{\omega}} = \boldsymbol{z}_{gyro} - \hat{\boldsymbol{b}}_{gyro}
\end{equation}
Where the hat symbol depicts the estimated quantities. The remaining states can be combined to formulate the EKF as
\begin{equation}
\boldsymbol{\zeta} = \left(\boldsymbol{x},\boldsymbol{\rho},\mathbf{\delta q}\right)
\end{equation}
Where $\boldsymbol{\rho} = \mathbf{R}^{-1}\dot{\boldsymbol{x}}$ is the velocity of the deputy, expressed in the chief-fixed reference frame and $\mathbf{\delta q}$ represents the error in the estimated attitude quaternion. Such formulation of attitude error instead of the absolute attitude offers robustness against some of the adverse effects arising from normalization constraint $\mathbf{q^Tq=1}$\cite{crassidis2011optimal}. The attitude error quaternion relates to the true attitude and estimated attitude by
\begin{equation}
\mathbf{\delta q} = \mathbf{q}\otimes\mathbf{\hat{q}}^{-1}
\end{equation}
At the start of each iteration, the error in attitude quaternion is reset to zero and inflated based on the current angular velocity estimate during the propagation step\cite{Markley}. Hence, the uncertainty in the estimated attitude is captured by the error quaternion and the need to track a separate covariance parameter for attitude estimate is no longer present\cite{OG_fusion}.  

\subsection{Prediction step}
For a given time instance, the predicted position and velocity of the module are computed by leveraging Equation~\eqref{eqn:coupled_EOM}, the current state, and control inputs. For the attitude error, the following prediction steps are followed\cite{crassidis2011optimal}, where $\boldsymbol{\delta \alpha}$ is the three-dimensional vector representing the attitude error.

\begin{equation}
\hat{\boldsymbol{\omega}} = \boldsymbol{z}_{gyro} - \hat{\boldsymbol{b}}_{gyro}
\end{equation}
\begin{equation}
    \boldsymbol{\delta\dot{\alpha}} = \hat{\boldsymbol{\omega}}
\end{equation}

\subsection{Measurement update}
For the measurement update, the attitude error is left unchanged\cite{crassidis2011optimal}, and the following range measurement equation is utilized to make a suitable correction in the predicted states computed earlier.
\begin{equation}
y = \lVert \boldsymbol{\rho}-\boldsymbol{\rho_A}\rVert
\end{equation}
Where $\boldsymbol{\rho_A}$ is the position vector of the anchor, and $y$ is the expected range. The Kalman gain is computed based on the difference between the measured and expected ranges, and the subsequent correction is done in the predicted states. Standard equations for the Kalman filter and Jacobian computations are not repeated here for brevity. As mentioned in the earlier section on the UWB sensor model, the range measurements often contain outliers which can significantly perturb the estimated states. Hence, Mahalanobis distance-based outlier rejection is implemented to ensure that the pose estimate is robust to such perturbations.

\subsection{Reset}
Once the measurement update is performed, the attitude quaternion is corrected based on the following quaternion kinematic equations and the attitude error vector,
\begin{equation}
    \hat{\textbf{q}}_k^+ = \hat{\textbf{q}}_k^- + \frac{1}{2}\Sigma\left(\hat{\textbf{q}}_k^-\right)\delta\hat{\boldsymbol{\alpha}}_k^+
\end{equation}
Where,
\begin{equation*}
    \Sigma\left(q\right) = \begin{bmatrix}
q_4I_3+[\![\varrho \times]\!]\\
-\varrho^{T}
\end{bmatrix}	
\end{equation*}
The attitude error vector is reset to zero once the estimated attitude quaternion is updated.

\begin{figure}[tb!]
\centerline{\includegraphics[width=0.9\textwidth]{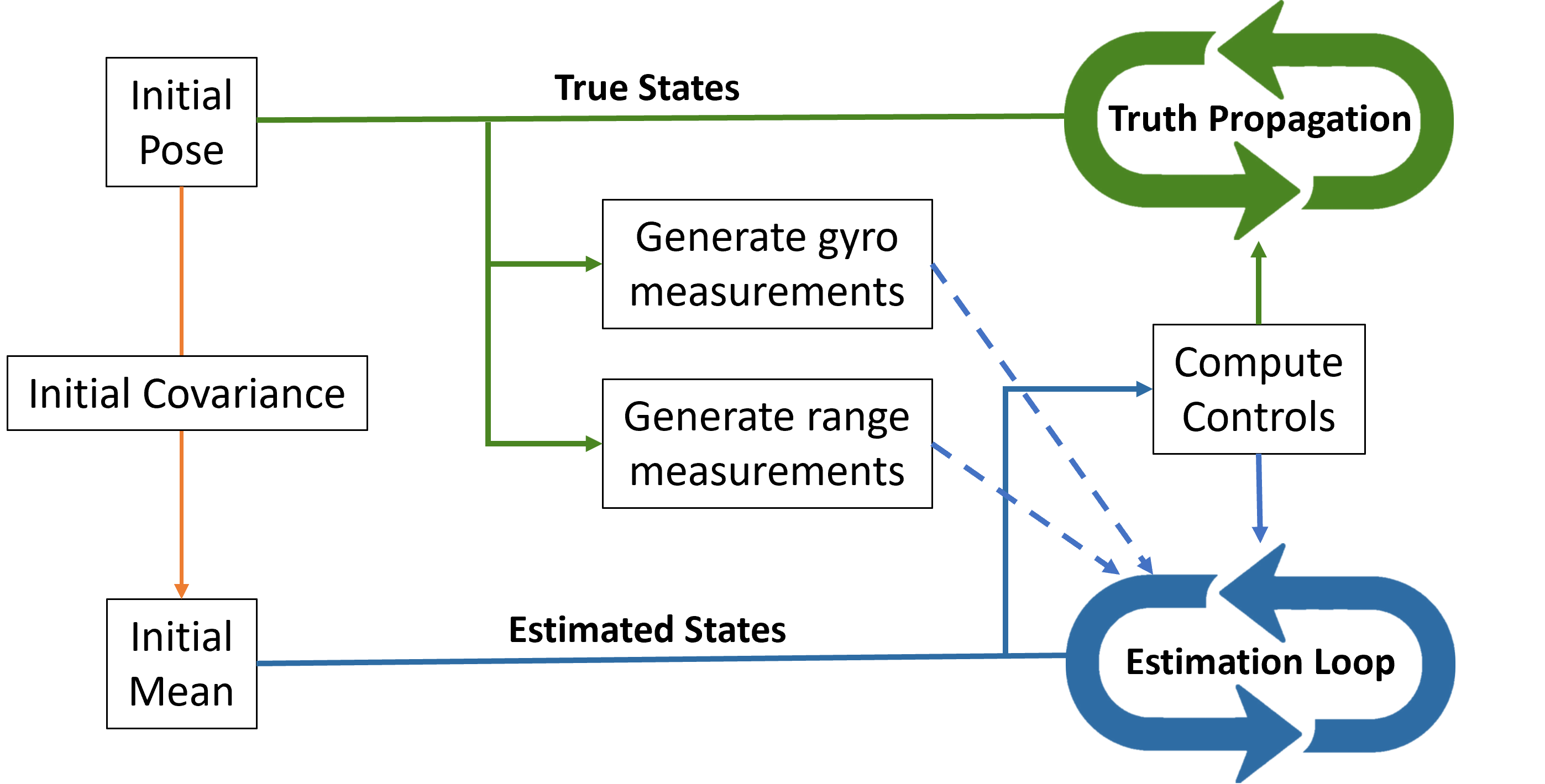}}
\caption{Simultaneous truth propagation and estimation with closed-loop control}
\label{fig:estm_loop}
\end{figure}

\section{Position controller}
The estimated pose is further leveraged to compute control inputs needed to drive TPODS to the desired relative pose. However, since the input states given to the closed-loop controller are their estimates, the final error in the pose is driven by a combination of errors in the pose estimate and the steady state error of the controller. The following subsections details the simulation architecture to compare true and estimated states under dynamic simulation.

\subsection{Simulation setup}
The simulation architecture employed is presented in Figure~\ref{fig:estm_loop}. Since the objective of the simulation is to assess the performance of the closed-loop position controller along with pose estimation, the true states are not known beforehand. This necessitates truth propagation along with the regular estimation loop. The initial mean is sampled from a known initial pose and covariance at the beginning of the simulation. The synthetic measurements for angular rates and range are generated from true states, and appropriate sensor models discussed earlier. A range from only one UWB anchor is available for a particular iteration. Hence, the specific anchor is picked from a uniform distribution of available anchors. In addition, $10\%$ of the total range measurements are corrupted by measurement noise, which is ten times worse than the sensor model, to represent outliers in range measurement.

Once the simulated measurements are available, the control inputs are computed based on the estimated states and desired set point. As presented in the next section, the controller's output is the desired force in the chief-attached reference frame. However, the thrusters impart forces in the deputy-attached body reference frame. Hence, the desired forces in the deputy-attached reference frame are computed from controller output using attitude estimate. Finally, the actual thrust commands are computed based on the thrusters' arrangement and TPODS geometry. The states of TPODS is propagated once the thrust commands are available. This process propagates true and estimated states while considering true and estimated current states as the initial condition respectively. As there is usually an offset between true and estimated attitude for the reasons discussed in the later section, the actual position departs from the desired position if the feedback is not employed via range measurements.

\begin{figure}[tb!]
    \centering
    \includegraphics[width=\textwidth]{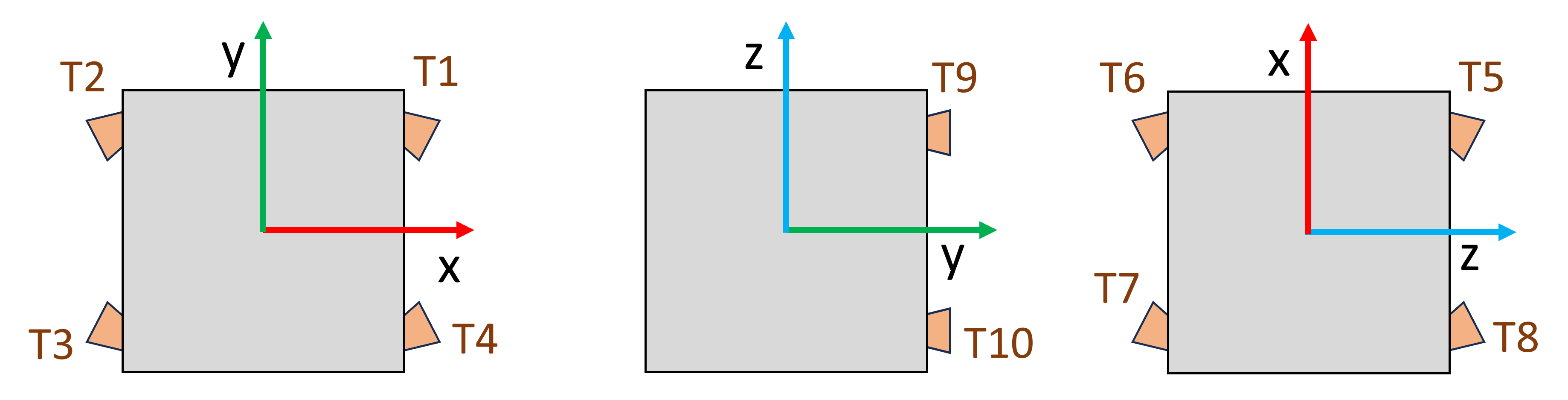}
    \caption{Thruster arrangement for TPODS}
    \label{fig:thrusters}
\end{figure}

\subsection{Controller structure}
A full state feedback-based Linear Quadratic Regulator(LQR) is implemented to drive the TPODS to a desired pose\cite{ogata2010modern}. Translation position and velocity, attitude, and rotational velocity are fed as inputs to the controller to compute desired forces in the chief-fixed reference frame. These desired force inputs are first converted to the desired forces in the deputy-attached reference frame and finally to the appropriate thruster commands. To compute thrust commands of each thruster from the desired forces, information about the arrangement of thrusters on the TPODS body is needed. For the scope of this analysis, arrangements of 10 thrusters, as shown in Figure~\ref{fig:thrusters} is considered. It ensures complete control of all six DOFs, with some added redundancy.
 
\section{Closed-loop performance}
\subsection{6-DOF Simulation}
\begin{figure}[htb!]
     \begin{subfigure}[b]{0.5\textwidth}
        \centering
         \includegraphics[width=\textwidth]{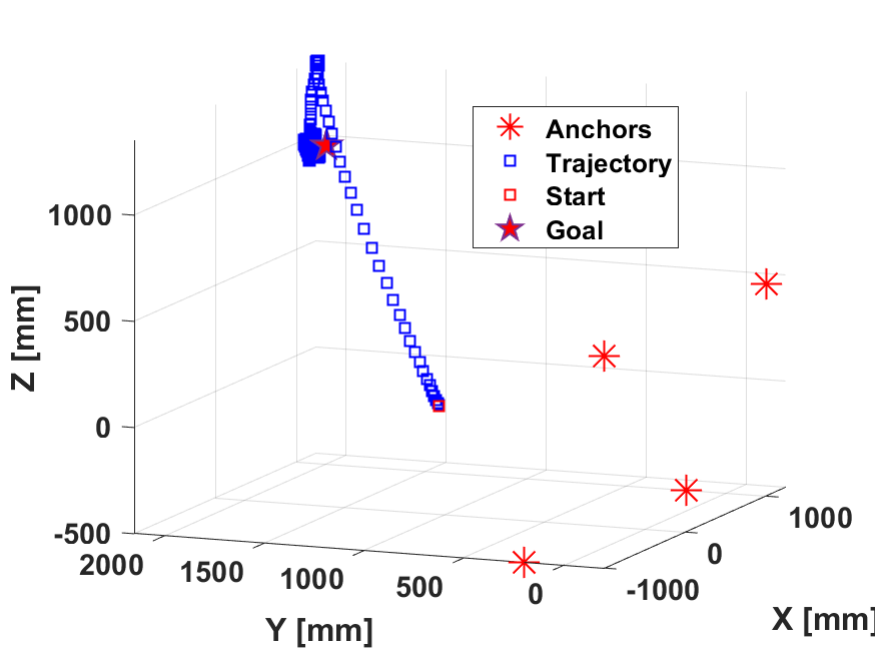}
	   \caption{}
		\label{fig:CL_traj}	
     \end{subfigure}
     \centering
     \begin{subfigure}[b]{0.49\textwidth}
        \centering
         \includegraphics[width=\textwidth]{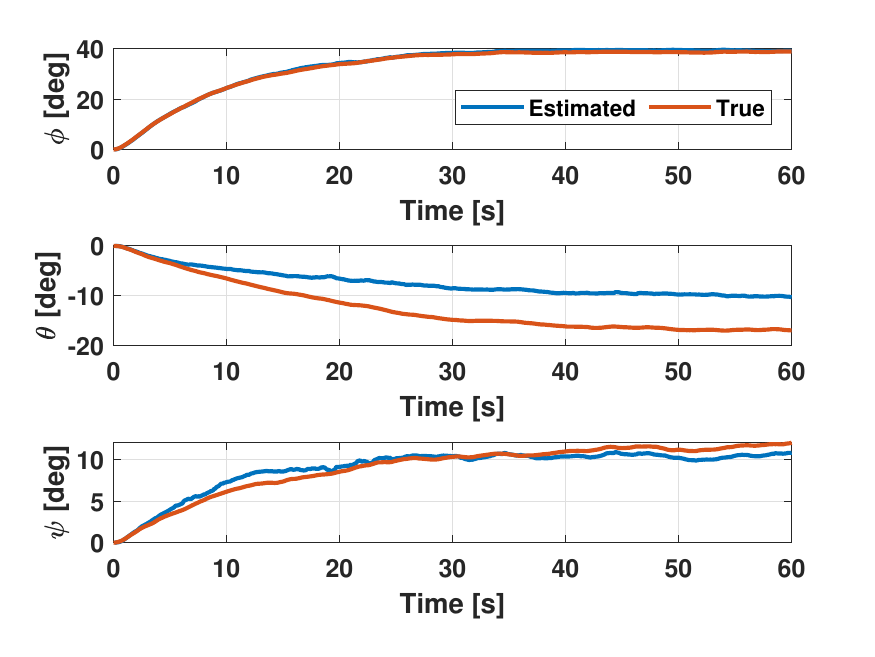}
	   \caption{}
		\label{fig:CL_attitude}
    \end{subfigure}
    \caption{Position and attitude history for closed-loop control simulation}
\end{figure}

The performance and tuning of the pose estimator are carried out by considering a couple of simulation scenarios and the subsequent generation of synthetic measurements. The TPODS module is commanded to reach the desired pose, as shown in Figure~\ref{fig:CL_traj} while honoring thrust saturation constraints. During the motion across the trajectory, the true states are recorded and utilized to generate synthetic measurements for angular velocity and relative range with fixed UWB anchor modules. The measurement models presented in an earlier section of this article are leveraged to introduce an appropriate measurement noise while generating simulated measurements, and the location of the fixed anchor modules is chosen based on the geometry of the mother-ship.

To assess the performance of the pose estimator for a combined translation and rotational motion, the TPODS module is commanded to reach a desired pose that demands a change in relative position and attitude. The guidance logic computes the desired attitude at each time instance based on the current position of the modules, and the control algorithm performs necessary corrections to ensure that the TPODS reaches the desired final pose. This type of motion ensures that both the translation and rotational modes are being excited during the duration of the motion.

\begin{figure}[b!]
    \centering
    \includegraphics[width=\textwidth]{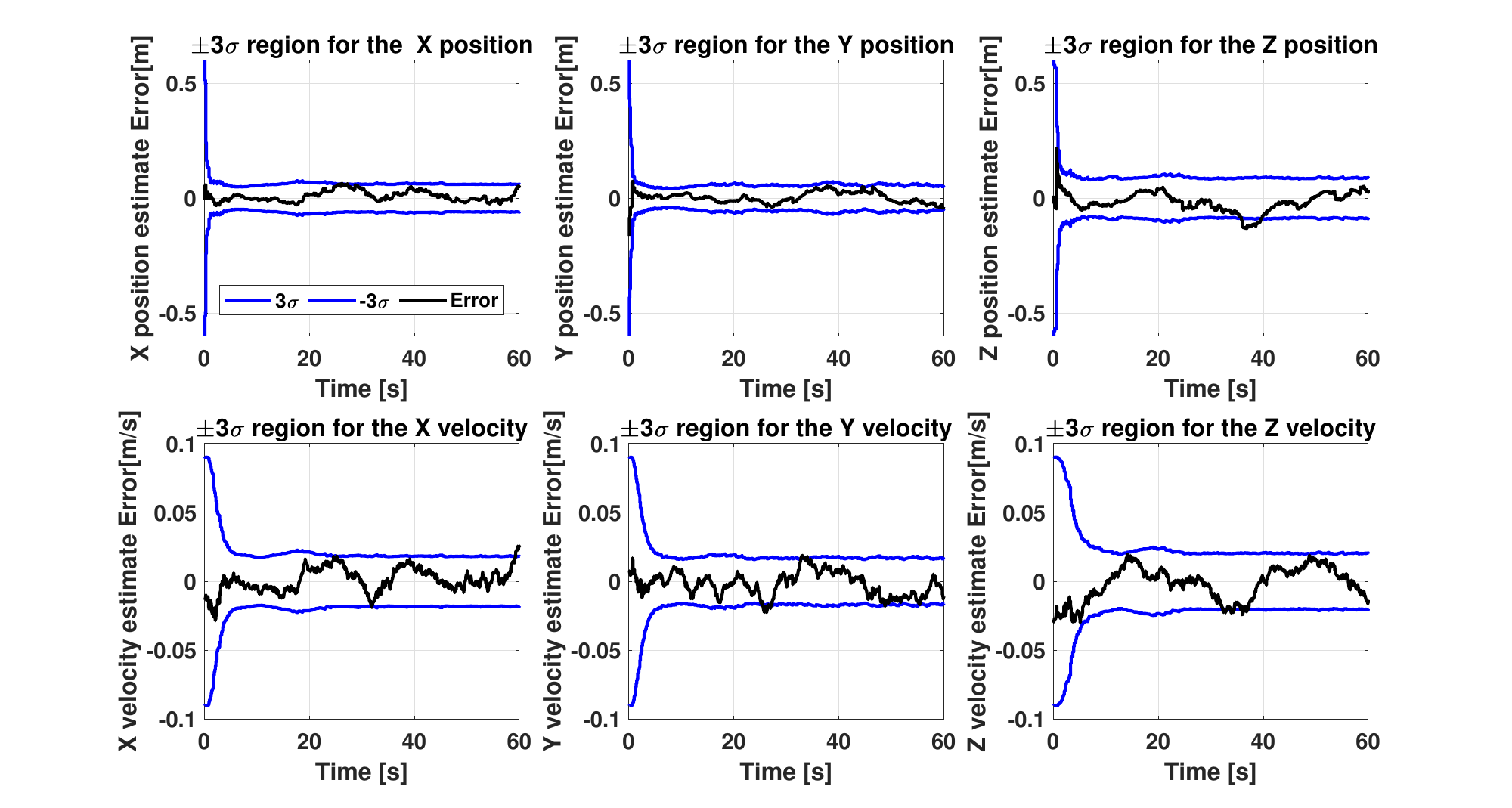}
    \caption{Estimation errors and respective variances}
    \label{fig:cl_3sigma}
\end{figure}

The performance of the pose estimator algorithm can be assessed by the plots of estimation error along with their respective variance histories as presented in Figure~\ref{fig:cl_3sigma}. It is evident from the figure that the estimation errors for all states stay within their respective $\pm3\sigma$ regions for all time, indicating that the filter parameters are well-tuned. The estimation error in position remains within the limit inferred from the accuracy of the range sensor. While the TPODS reaches the commanded 3-D position with acceptable steady state error, the estimated attitude significantly differs from the true value as shown in Figure~\ref{fig:CL_attitude}. This behavior is due to the absence of independent attitude measurements. Since measurements from the gyro are corrupted with white noise, the attitude data obtained by integrating the angular rates will deviate from the true attitude. For a typical spacecraft application, measurements from startrackers are employed periodically to correct the attitude estimate and respective gyro bias parameters. However, any deviation in the position is detected by the range sensors, and even though the attitude estimate is not aligned with the true attitude, the TPODS module still reaches the commanded position. Hence, the omission of the startracker is primarily driven by the mission's objective. Once the chaser TPODS reaches near the target, more accurate vision-based pose estimation can be leveraged for precise proximity operations\cite{TPODS_ICRA}. 	 

\section{Experimental Validation}
The effectiveness of the pose estimator, closed-loop position controller, and guidance logic are assessed via two scenarios. The closed-loop position controller is validated using a test setup that enables a planar motion of TPODS within the convex hull of four UWB anchors\cite{TPODS_ICRA}. For the second scenario, the UWB tag is mounted with an offset on a turntable, ensuring a kinematic coupling. The IMU is mounted at a center, and respective measurements and ground truth are recorded for further analysis.

\subsection{Closed-loop Position Control for Planar motion}
\subsubsection{Experimental Setup :}
\begin{figure}[b!]
     \begin{subfigure}[b]{0.51\textwidth}
        \centering
         \includegraphics[width=\textwidth]{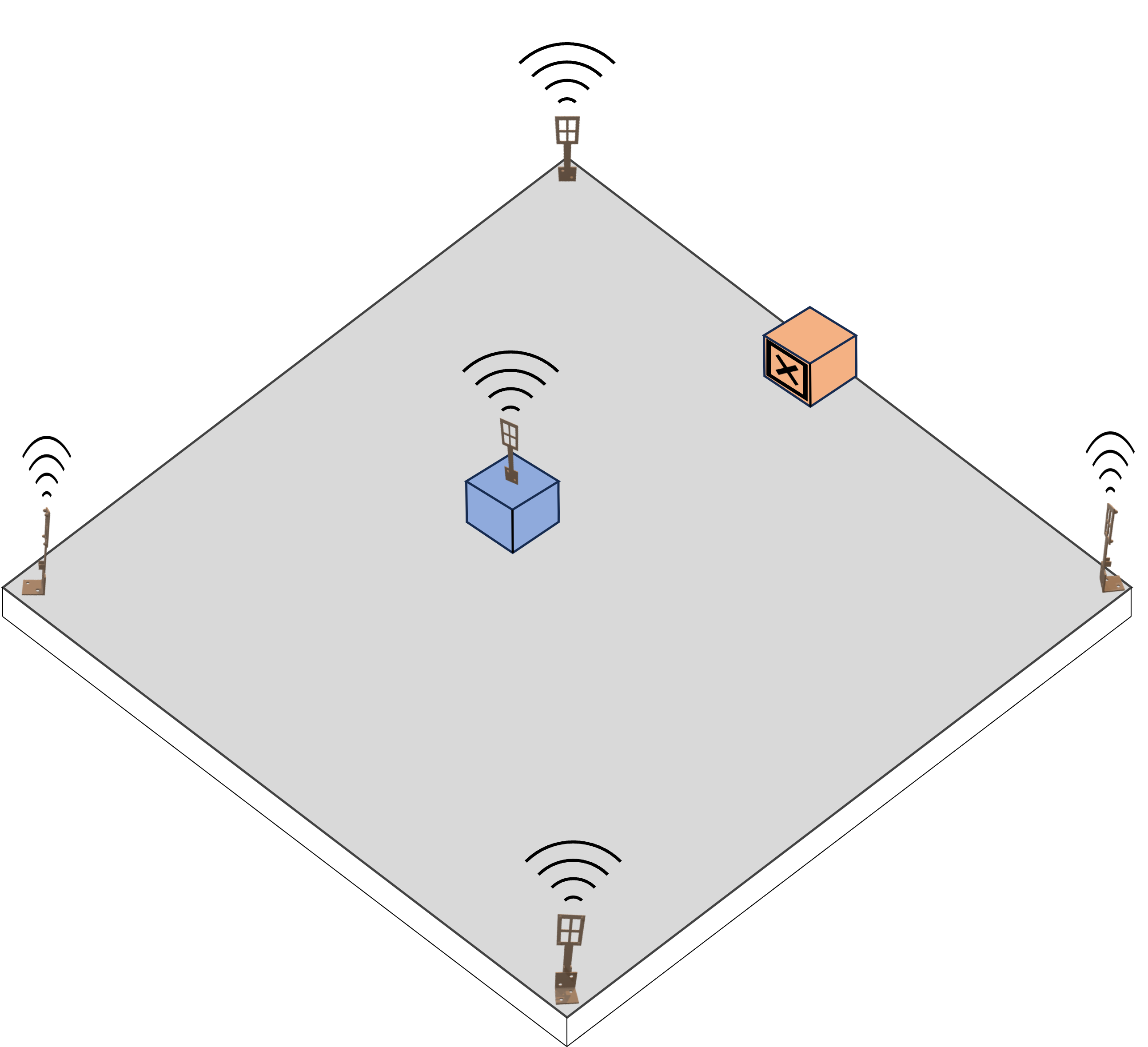}
     \end{subfigure}
     \centering
     \begin{subfigure}[b]{0.48\textwidth}
        \centering
         \includegraphics[width=\textwidth]{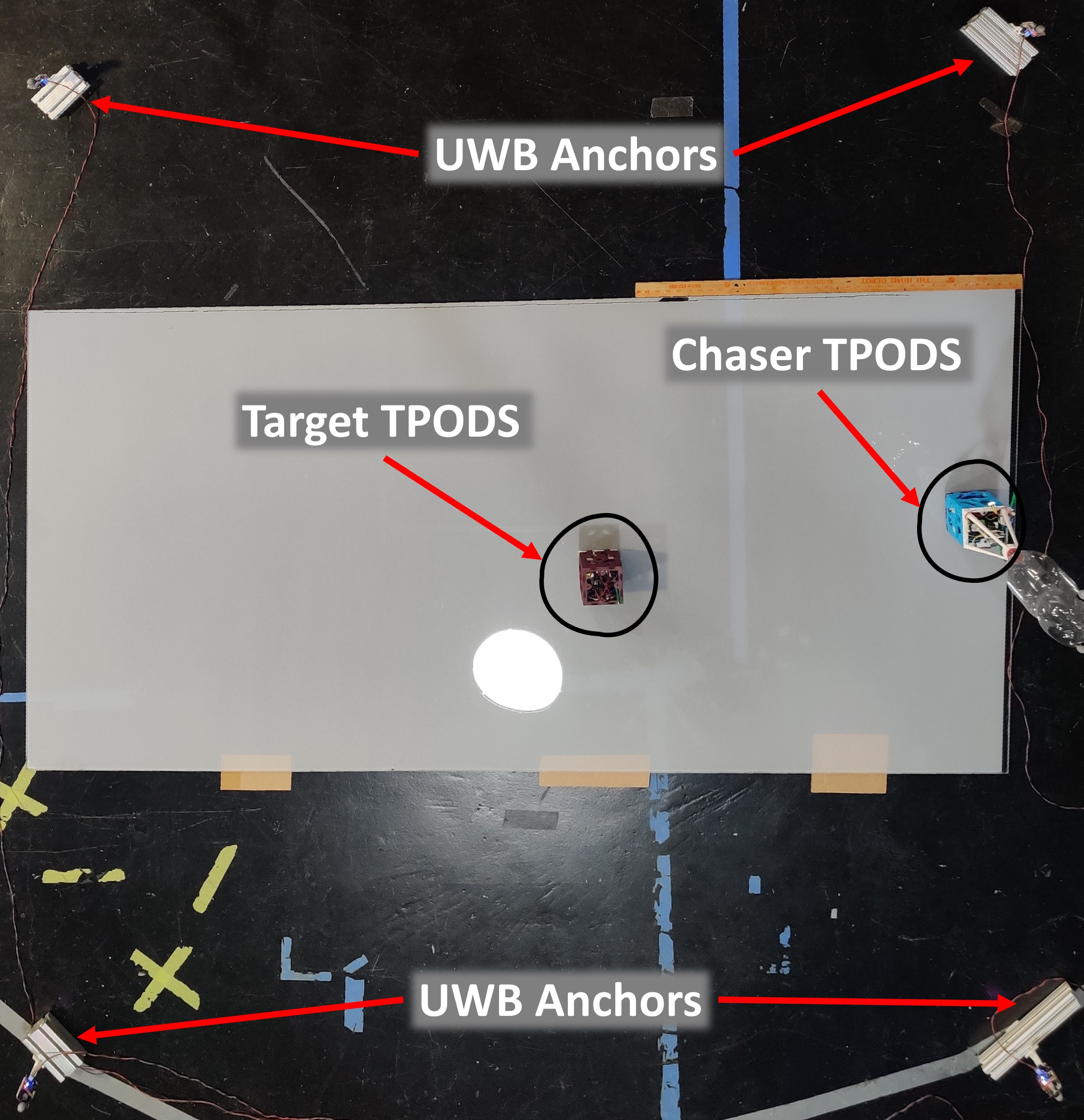}
    \end{subfigure}
    \caption{Experimental setup to validate pose estimator and closed-loop position control in 3-DOF}
    \label{fig:exp_setup_dock}
\end{figure}
To validate the performance of the pose estimator, a 3-DOF experimental setup of Figure~\ref{fig:exp_setup_dock} is leveraged. A TPODS module is equipped with VectoNav VN-100 Inertial Measurement Unit (IMU) and is polled to pass angular rates and acceleration data every $10ms$. The module also consists of a Loco positioning node from BitCraze AB, which can determine the relative range from fixed UWB anchors. Since the module is constrained to move in a plane, four UWB anchors are considered for the relative range measurements. The node mounted on the moving TPODS module polls each anchor randomly and provides relative range data to the pose estimator.

Teensy 4.1 assumes the role of central compute board and ensures measurement acquisition from various sensors. The measurements and their respective timestamps are stored on the SD card for offline filter analysis. The LASR lab is also equipped with a motion capture system that can provide pose measurements of an object with positional accuracy in the order of a few millimeters and attitude accuracy in the order of a few degrees. The motion capture system is utilized to log pose values during the motion of TPODS, and recorded data is treated as ground truth to access the performance of the pose estimator developed in earlier sections of this paper. 
\subsubsection{Performance analysis :}

During the validation experiment, the chaser TPODS is commanded to move toward a fixed target. The pose estimator runs onboard Teensy 4.1 and utilizes range and IMU measurements to achieve a good initial position and orientation estimate. Once the lock on the initial pose is established, the LOS guidance computes the correction in orientation needed to point towards the target and the forward velocity based on the distance from the target. The control allocation scheme is designed to give higher priority to the attitude correction commands than the translation commands. Such prioritization is needed as the translation and rotation demands are often fulfilled by firing the same thrusters. It is observed that for various initial positions and orientations, the chaser TPODS can align its heading towards the target and reach within the expected region around the fixed target, demonstrating the effectiveness of onboard pose estimator and closed-loop position control\cite{TPODS_ICRA}.

\subsection{Open-loop pure rotation in a plane}
\subsubsection{Experimental Setup :}
\begin{figure}[b!]
     \begin{subfigure}[b]{0.46\textwidth}
        \centering
         \includegraphics[width=\textwidth]{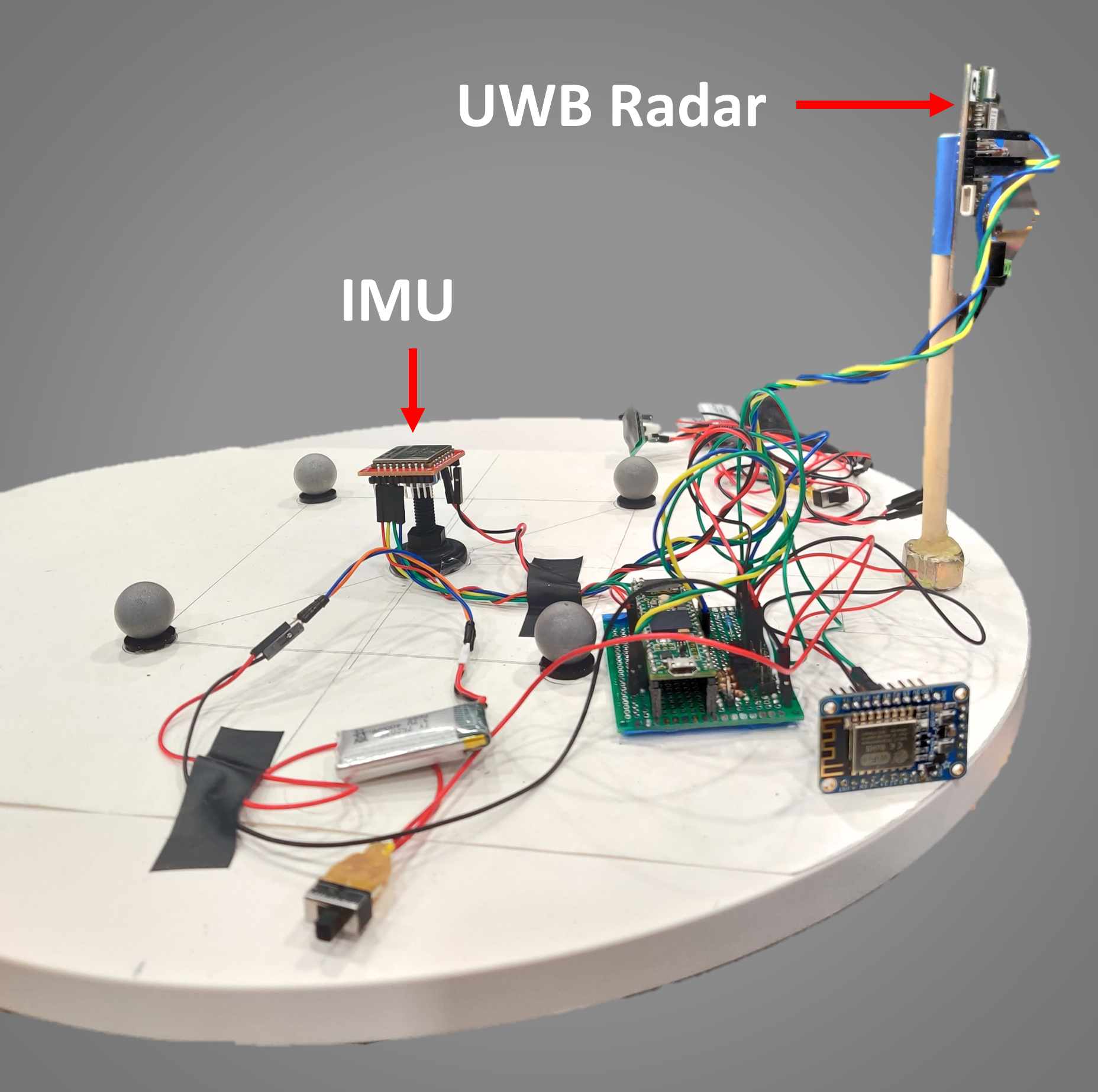}
     \end{subfigure}
     \centering
     \begin{subfigure}[b]{0.53\textwidth}
        \centering
         \includegraphics[width=\textwidth]{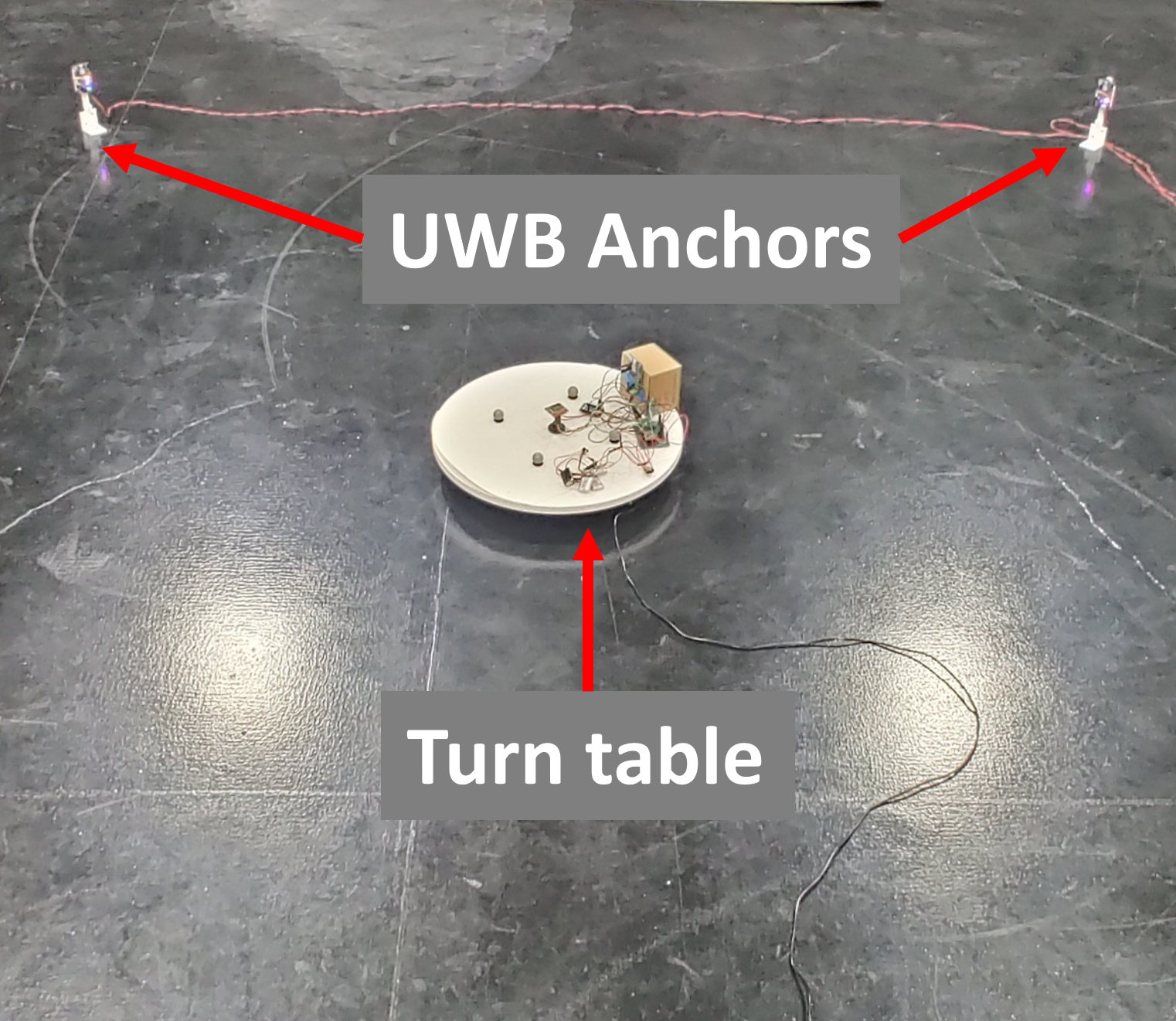}
    \end{subfigure}
    \caption{Experimental setup to validate pose estimator with kinematic coupling}
    \label{fig:exp_setup_turn}
\end{figure}

As presented in Figure~\ref{fig:exp_setup_turn}, IMU and UWB range sensors are arranged on a turn table to ensure kinematic coupling between translation and rotational motion. The IMU is mounted at a carefully marked center of rotation to ensure that the measurements from the gyro are induced by rotation only. However, the UWB is mounted at an offset with the center of rotation, resulting in kinematic coupling. The Teensy 4.1 compute board serves a similar role as explained earlier and logs necessary parameters and measurements for further offline analysis. The entire turn table assembly is placed within a convex hull of four UWB anchors. Hence, when the turn table is commanded to rotate, the respective ranges show oscillatory motion, even though the turn table does not undergo any translation motion. The turn table is also equipped with reflective markers to enable the logging of ground truth via a motion capture system.
\subsubsection{Performance analysis :}

\begin{figure}[b!]
     \begin{subfigure}[b]{0.495\textwidth}
        \centering
         \includegraphics[width=\textwidth]{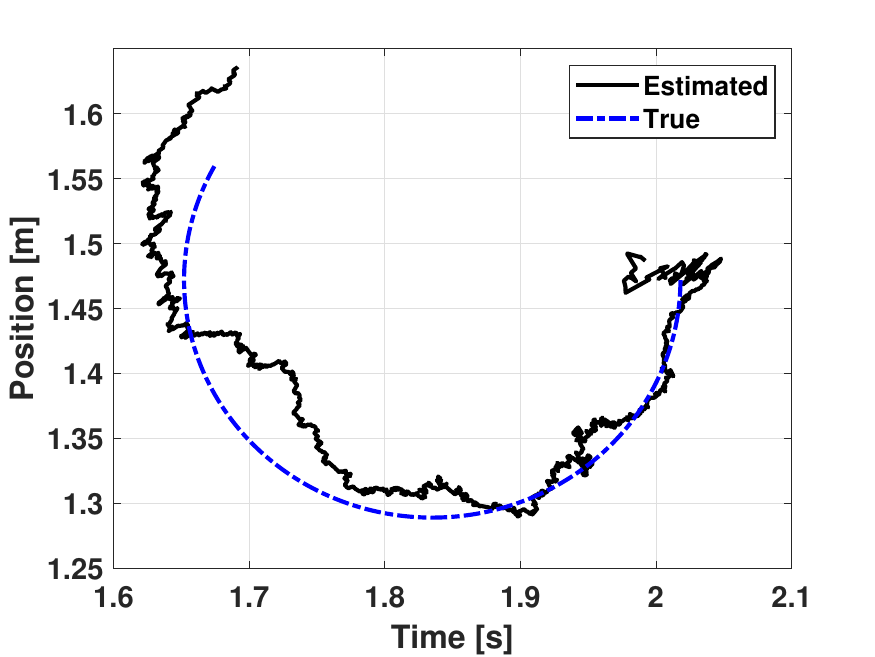}
     \end{subfigure}
     \centering
     \begin{subfigure}[b]{0.495\textwidth}
        \centering
         \includegraphics[width=\textwidth]{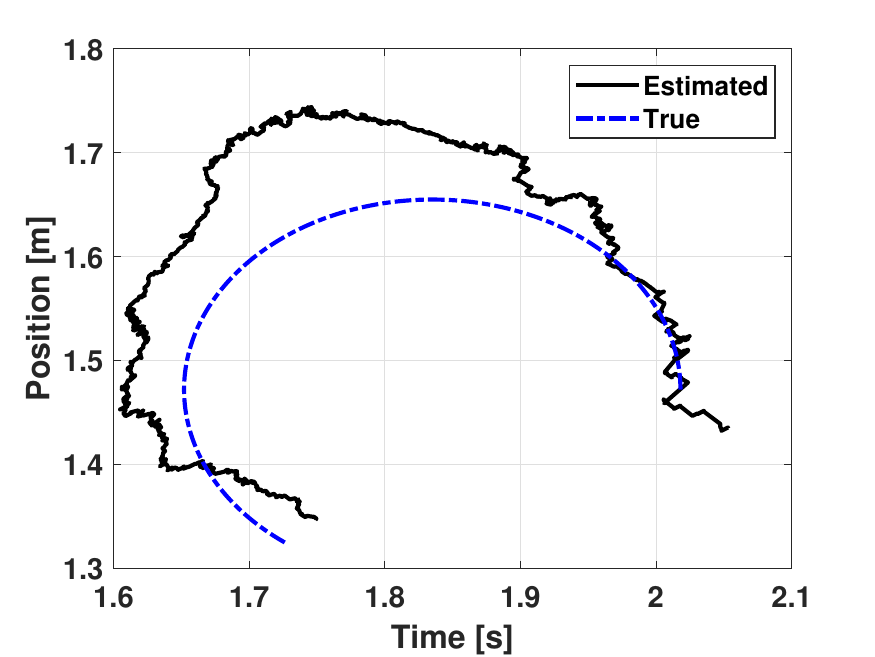}
    \end{subfigure}
    \caption{Estimated pose from the pose estimator and ground truth}
    \label{fig:pos_turn}
\end{figure}

The recorded measurements are fed to the pose estimation algorithm, carefully configured to capture the kinematic coupling due to the actual arrangement of IMU and UWB range sensors. It can be concluded from Figure~\ref{fig:pos_turn} that for both clockwise and counterclockwise rotation scenarios, the pose estimator can predict the general nature of the motion of the UWB tag. A detailed analysis of the pose estimator reveals that the number of rejected range samples is not adversely affected due to the kinematic coupling when compared to the first scenario where the UWB tag is mounted on axes of rotation of the module. This is compelling evidence that the pose estimator effectively captures the effects of kinematic coupling.

\section{Future work}
Although the pose estimation algorithm predicts the general nature of the motion well for the pure rotation in the planar case, the estimator's performance needs to be examined for free rotation in all three axes of rotation. Since the gain characteristics of UWB sensors used for this analysis vary based on the plane of motion\cite{mymanual,sorgel2005influence}, the estimator's performance is expected to deteriorate for some regions of motion. A pose estimator algorithm aware of variable gain characteristics can be an avenue to deal with such shortcumings effectively.

\section{Acknowledgment}
This work is supported by the Air Force Office of Scientific Research (AFOSR), as a part of the SURI on OSAM project “Breaking the Launch Once Use Once Paradigm” (Grant No: FA9550-22-1-0093). Program monitors for the AFOSR SURI on OSAM, Dr. Frederick Leve of AFOSR, and Mr. Matthew Cleal of AFRL are gratefully acknowledged for their watchful guidance. Prof. Howie Choset of CMU, Mr. Andy Kwas of Northrop Grumman Space Systems and Prof. Rafael Fierro of UNM are acknowledged for their motivation, technical support, and discussions.

\bibliographystyle{AAS_publication}   % Number the references.
\bibliography{references}   % Use references.bib to resolve the labels.
\end{document}